\documentclass[twocolumn,preprintnumbers,aps,showpacs,amsmath,amssymb]{revtex4}
\usepackage{graphicx}
\usepackage{bm}
\usepackage{float}

\newcommand{\mL}{{\mathcal{L}}}
\newcommand{\mE}{{\mathcal{E}}}
\newcommand{\mB}{{\mathcal{B}}}

\begin{document}

%\preprint{arXiv:1810.xxxx [hep-ph]}

\title{Virtual axion-like particle complement to Euler-Heisenberg-Schwinger action}

\author{Stefan Evans}
\author{Johann Rafelski}
\affiliation{Department of Physics, The University of Arizona, Tucson, AZ 85721, USA}
%\date{\today}
\date{October 15, 2018}

\begin{abstract}
We modify action in an external electromagnetic field to include effects of virtual axion-like particle (ALP) excitations. A measurable addition to QED-Euler-Heisenberg-Schwinger (EHS) action is obtained and incorporated into experimental constraints placed on ALP mass and coupling to two photons. The regime of these constraints in which the ALP vacuum effect surpasses the EHS effect is characterized. We show that probing of the virtual vacuum effect offers an alternative method in search for physics related to ALPs.
\end{abstract}

\pacs{14.80.Va,11.15.Tk,11.40.Ha,13.40.-f}
% 14.80.Va Axions and other Nambu-Goldstone bosons (Majorons, familons, etc.) 
% 11.15.Tk Other nonperturbative techniques
% 11.40.Ha Partially conserved axial-vector currents 
% 13.40.-f Electromagnetic processes and properties
%NOT  13.40.Em Electric and magnetic moments 
%NOT 13.40.Dk Electromagnetic mass differences 
%NOT 12.20.-m Quantum electrodynamics

\maketitle

\section{Introduction}

We show that a pseudoscalar coupling between two photons and axion-like particles (ALP) produces a nonlinear in EM field addition to the QED vacuum effect akin to a specific term in the Euler-Heisenberg-Schwinger (EHS) action~\cite{Heisenberg:1935qt,Schwinger:1951nm}. We report on this additional effect and the conditions required for it to be of or above in magnitude to the QED vacuum fluctuation  EHS result.  

The Adler-Bell-Jackiw anomaly is often used in study of pseudoscalar decay process $\pi^0\to\gamma\gamma$~\cite{Adler:1969gk,Bell:1969ts}. This was followed by proposed Axion-Like-Particle (ALP) $\leftrightarrow\gamma\gamma$ processes. Searches for these processes focus on possible   flux of on-mass shell ALP particles from astronomical sources~\cite{Jaeckel:2010ni,Tanabashi:2018oca}. Another consequence of the ALP$\leftrightarrow\gamma\gamma$ interaction is a possible modification of the vacuum by virtual fluctuations which we describe here.

\begin{figure}[H]
\center
\includegraphics[width=.85\columnwidth]{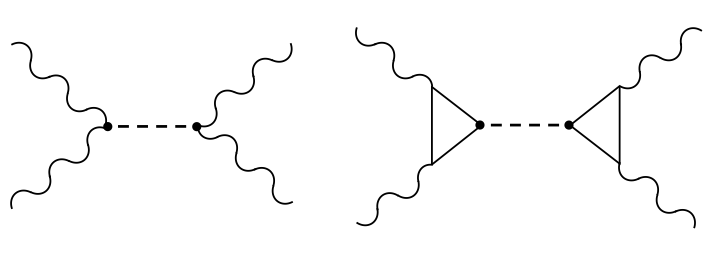}
\caption{\label{fig1} EHS complements: vacuum fluctuation contributions fourth order in the photon field. Left: four photon interactions via virtual ALP;  and Right via PCAC pion anomaly. } 
\end{figure}

The virtual EHS effect of the electron loop on electromagnetic processes has motivated measurement of vacuum birefringence in strong magnetic fields~\cite{Zavattini:2005tm,Zavattini:2007ee,DellaValle:2015xxa}. We complement this effective QED-EHS action including effects on vacuum structure by two other effects, the virtual ALP excitations and pions -- the ALP couples to two photons with point particle form factor while the pion via quark triangle loop, see Figure~\ref{fig1}. 

The ALP supplement to EHS action is obtained by rescaling ALP fields to include photon-loop corrections. This produces a fourth-order in EM field addition to EHS action, a consequence of ALP fluctuations introducing an extra diagram contributing to vacuum polarization shown in the left-hand-side of Figure~\ref{fig1}. 

%%%%%%%%%%%%
%
%
%
%
%%%%%%%%%%%% 

The strength of the ALP fluctuation supplement to EHS action is characterized by ALP-$\gamma\gamma$ coupling involving the ratio of coupling and mass $G_\mathrm{A}/m_\mathrm{A}$. Observations of astrophysical ALP sources have provided constraints on $G_\mathrm{A}$, $m_\mathrm{A}$: see~\cite{Jaeckel:2010ni,Tanabashi:2018oca} and references therein. Unlike in the case of axions, for ALPs $G_\mathrm{A}/m_\mathrm{A}$ is not fixed. While the fixed value of pseudoscalar-$\gamma\gamma$ coupling to mass ratio (or product of mass and decay constant) excludes significant additions to EHS action in the case of axions and pions, the ALP contribution does not have this restriction.

In the following we evaluate the size of the ALP contribution complement to the  EHS effective action in section~\ref{supplement}. We describe the experimental constraints in section~\ref{evaluated}, where the domain of $G_\mathrm{A}/m_\mathrm{A}$ is shown in which the ALP contribution surpasses the corresponding contribution from the EHS action.

%%%%%%%%%%%%%%%%%%%%%%%%%%%%%%%%%%%%%%%%%%%%%%%%%%%
\section{Pseudoscalar Coupling input to effective EHS Action}
\label{supplement}

\subsection{ALP complement to  EHS}
\label{ALPfirst}

We consider the effective QED action using EM field invariants
\begin{equation}
S=\frac12(\mE^2-\mB^2)\;,\qquad
P=\mE\cdot\mB
\;,
\end{equation} 
in the respective terms 
\begin{equation}
\label{total}
\mL=S+\mL_{EHS}+\mL_\varphi+\mL_{int} 
\;.
\end{equation}
We see first the Maxwell action, complemented by renormalized effective EHS action, since in the interaction term bare $(e_0\mE_0)(e_0\mB_0)$ is equivalent to renormalized $(e\mE)(e\mB)$:
\begin{align}
\label{EH_genEq}
\mL_{EHS}=
m_e^4f\Big(\frac S{m_e^4},\frac {P^2}{m_e^8}\Big) 
\;,
\end{align}
where $m_e$ is electron mass and the function $f$ is well known \cite{Heisenberg:1935qt,Schwinger:1951nm}. Note that for reason of parity conservation by the QED vacuum, a $P$-term must have even powers in Eq.\,(\ref{EH_genEq}). The low field expansion generates the well known term
\begin{align}
\label{perturb}
\mL_{EHS}^{(1)}=\frac{e^4}{(4\pi)^2}\frac{2}{45m_e^4}\Big(4S^2+7P^2\Big) 
\;,
\end{align}
where superscript $(1)$ denotes action up to the leading nonlinear EM contribution.

The two supplemental  and  here relevant ALP terms in Eq.\,(\ref{total}) include  the pseudoscalar mass contribution
\begin{equation}
\mL_\varphi=-\frac{m_\mathrm{A}^2}2 \varphi^2 
\;,
\end{equation}
where $\varphi$ is the ALP  field. We neglect the kinetic energy term, as we are interested in the infrared limit of vacuum fluctuation contribution.  For the effective ALP with two photon interaction term~\cite{Jaeckel:2010ni,Tanabashi:2018oca}
\begin{equation}
\label{Lint}
\mL_{int}= \pm G_\mathrm{A}\varphi P
\;,
\end{equation}
where ALP to two photon coupling $g_A[\mathrm{GeV}^{-1}]$ is constrained by experimental observation. Note that the sign in the effective action Eq.\,(\ref{Lint}) requires further consideration: its value will not appear in our computations which are quadratic in this term, thus we choose in further considerations the positive sign.

The total action we thus consider comprises three contributions:  EHS Eq.\,(\ref{perturb}) and the two ALP terms
\begin{align}
\label{TotalV2}
\mL^{(1)}=&\,
S-\frac{m_\mathrm{A}^2}2 \varphi^2+ G_\mathrm{A}\varphi P
\nonumber \\
+&\frac{e^4}{(4\pi)^2}\frac{2}{45m_e^4}\Big(4S^2+7P^2\Big) 
\;.
\end{align}

The ALP degrees of freedom can be \lq rotated\rq\, to diagonalize the ALP-$\gamma\gamma$ interaction. Completing the square,
\begin{align}
-\frac{m_\mathrm{A}^2}2 \varphi^2+ G_\mathrm{A}\varphi P
=&\,-\frac{m_\mathrm{A}^2}2 \Big(\varphi - G_\mathrm{A}\frac{ P}{m_\mathrm{A}^2}\Big)^2
+\frac{G_\mathrm{A}^2}{2m_\mathrm{A}^2}P^2
\nonumber \\
=&\,
-\frac{m_\mathrm{A}^2}2 \tilde\varphi^2
+\frac{G_\mathrm{A}^2}{2m_\mathrm{A}^2}P^2
\;,
\label{TotalV3}
\end{align}
where the ALP-field is configuration mixed  with electromagnetic $\gamma\gamma$-contribution
\begin{align}
\label{rescale}
 \tilde\varphi=\varphi- G_\mathrm{A}\frac{ P}{m_\mathrm{A}^2}
\;.
\end{align}
After configuration mixing there is an additional contribution, the last term in Eq.\,(\ref{TotalV3}), 
which like the considered term of  EHS action (second term in Eq.\,(\ref{perturb})) is proportional to $P^2$  but only depends on ALP properties and electron charge, and is thus independent of electron mass. 

The total action up to 4th order in EM fields is
\begin{align}
\mL=&\,
S-\frac{m_\mathrm{A}^2}2 \tilde\varphi^2+\tilde\mL_{EHS}
\;,
\end{align}
where the first term is the QED Maxwell field action, followed by ALP mass contribution and 
\begin{align}
\tilde\mL_{EHS}=&\,\mL_{EHS}+\frac{G_\mathrm{A}^2}{2m_\mathrm{A}^2}P^2
\;,
\end{align}
independent of the sign of the ALP interaction term in Eq.\,(\ref{Lint}), as mentioned earlier. Using the lowest order term in EHS action (Eq.\,(\ref{perturb})),
\begin{align}
\label{redoforpion}
\tilde\mL_{EHS}^{(1)}=&\,
\frac{e^4}{(4\pi)^2}\frac{2}{45m_e^4}\Big(4S^2+7P^2\Big)
+\frac{G_\mathrm{A}^2}{2m_\mathrm{A}^2}P^2
\;.
\end{align}
This is equivalent to modification of the $P^2$ coefficient in Eq.\,(\ref{perturb}) by
\begin{align}
\label{coeff}
P^2
\to&\, P^2\Big(1+\frac{45m_e^4}{28\alpha^2}\frac{G_\mathrm{A}^2}{m_\mathrm{A}^2}\Big)
\;,
\end{align}
where $\alpha=e^2/4\pi$.

In order for the ALP effect to be equal or larger than the QED effect, the ratio of ALP-$\gamma\gamma$ coupling to mass $m_\mathrm{A}$ amounts to
\begin{align}
\label{forplot}
\frac{G_\mathrm{A}^2}{m_\mathrm{A}^2}\geq\frac{28\alpha^2}{45m_e^4}
\;.
\end{align}
This condition will be used to characterize presence of a virtual pseudoscalar effect on electromagnetism comparable to electron loop effects, in context of constraints from astronomical observation and vacuum birefringence experiment. We first discuss the differences between ALPs, axions and pions in context of Eq.\,(\ref{forplot}).

%%%%%%%%%%%%%%%%%%%%%%%%%%%%%%%
\subsection{Pion and axion}
\label{pionaxion}

Evaluation of the $P^2$ rescaling in Eq.\,(\ref{coeff}) is repeated for the pion, a result of supplementary diagram to QED vacuum polarization shown in the right-hand-side of Figure~\ref{fig1}. Eq.\,(\ref{Lint}) is replaced with
\begin{equation}
\mL_{int}= \frac {\alpha}{2\pi f_\pi}\varphi P
\;,
\end{equation}
where pion decay constant $f_\pi=93\,$MeV~\cite{Huang:1982ik}. This is reminiscent of Schwinger's result characterizing pion decay, which was missing a phase factor enforcing gauge invariance, summarized and reconciled with PCAC in~\cite{Zumino:1969mz,Treiman:1986ep,tHooft:2005hbu}.

%%%%%%%%%%%% Figure grossly modified JR
%
%
\begin{figure*}
\center
\includegraphics[width=1.8\columnwidth]{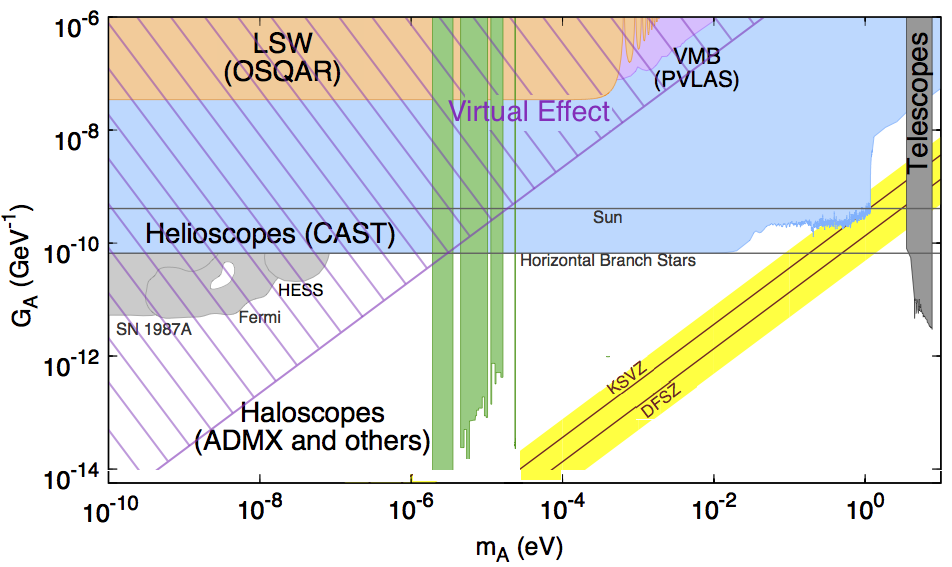}
\caption{\label{fig2} We adapt Figure 111.1 from Ref.\cite{Tanabashi:2018oca}, adding purple lines to denote a new region labeled \lq virtual effect\rq, where according to Eq.\,(\ref{forplot3}) the ALP supplement to effective action surpasses the first nonlinear EHS effect.} 
\end{figure*}
%
%
%%%%%%%%%%%% 

Repeating steps in section~\ref{ALPfirst}, with $\varphi$ now denoting a pion field and $m_A\to m_\pi\sim 135\,$MeV,
\begin{align}
P^2
\to&\, P^2\Big(1+\frac{45m_e^4}{112\pi^2}\frac{1}{f_\pi^2m_\pi^2}\Big)
\nonumber \\
=&\,
P^2(1+\mathcal O(10^{-10}))
\;,
\end{align}
producing a negligible addition to EHS action. Using instead an axion-field, the result is still characterized by a product $f_am_a$ on the order of $f_\pi m_\pi$~\cite{Jaeckel:2010ni,Tanabashi:2018oca}: the virtual axion effect does not produce a significant contribution to $\tilde \mL_{EHS}$ either.
 
Even though the result for pions is very small, we note that the decay constant $f_\pi$ is measured  in timelike kinematic domain with pion momentum $Q_\pi^2\to m_\pi^2$. Whether application of $f_\pi=93\,$MeV to the infrared domain $Q_\pi^2\to 0$ of external EM fields is valid remains an open  question. While this issue of kinematic domains awaits resolution, we only consider vacuum fluctuations of ALPs and not pions nor axions. We proceed to experimental constraints for ALPs.

%%%%%%%%%%%%%%%%%%%%%%%%%%%%%%%%%%%%%%%%%%
\section{Evaluated ALP vacuum fluctuation effect}
\label{evaluated}

\subsection{Constraints on $G_\mathrm{A}$ and $m_\mathrm{A}$}
\label{constraints}

We wish to add our result  Eq.\,(\ref{forplot}) into constraints on $G_\mathrm{A}$ and $m_\mathrm{A}$, based on results provided in  Figure 111.1 from~\cite{Tanabashi:2018oca}. All these constraints are based on real axions, with the exception of virtual effects probed by the vacuum birefringence measurements carried out by the PVLAS~\cite{Zavattini:2005tm,Zavattini:2007ee,DellaValle:2015xxa} collaboration. Our addition is  a virtual effect, a supplement to EHS effective action. 

For clarity we describe in detail how the new domain arises:
Taking the square root of Eq.\,(\ref{forplot}) and writing units explicitly, 
\begin{align}
\label{forplot2}
\frac{G_\mathrm{A}[\mathrm{GeV}^{-1}]}{m_\mathrm{A}[\mathrm{eV}]}\geq\sqrt{\frac{28}{45}}\alpha\frac1{(0.511\,\mathrm{MeV})^2}
\;.
\end{align}
We write
\begin{align}
G_\mathrm{A}[\mathrm{GeV}^{-1}]=\frac{10^y}{\mathrm{GeV}}\;,\qquad
m_\mathrm{A}[\mathrm{eV}]=10^x\, \mathrm{eV}
\;,
\end{align}
where the coefficients $y$, $x$ serve as variables $\log_{10}G_\mathrm{A}[\mathrm{GeV}^{-1}]$, $\log_{10}m_\mathrm{A}[\mathrm{eV}]$ in Figure~\ref{fig2}.
Eq.\,(\ref{forplot2}) becomes
\begin{align}
\frac{10^{y-x-9}}{(\mathrm{eV})^2}\geq\sqrt{\frac{28}{45}}\alpha\frac{10^{-12}}{(0.511)^2(\mathrm{eV})^2}
=\frac{2.21\cdot 10^{-14}}{(\mathrm{eV})^2}
\;.
\end{align}
Taking $\log_{10}$ and keeping only $y$ on the left hand side, 
\begin{align}
\label{forplot3}
y\geq x+ \log_{10}2.21\cdot 10^{-5}
= x-4.66
\;.
\end{align}
With this we define boundary at which ALP and QED effects equal:
\begin{align}
\label{forplot4}
y_{QED}= x_{QED}-4.66
\quad
\to \quad
\frac{G_\mathrm{A(QED)}}{m_\mathrm{A(QED)}}=\frac{10^{-4.66}}{(\mathrm{GeV})(\mathrm{eV})}
\;.
\end{align}

In Figure~\ref{fig2}, the region of Figure 111.1 of~\cite{Tanabashi:2018oca} where $y,x$ satisfy Eq.\,(\ref{forplot3}) is shaded. The boundary of the shaded region, at which the ALP and QED effects are equal is given by Eq.\,(\ref{forplot4}). This boundary is parallel to the KSVZ and DFSZ models for the axion shown in the figure, which have fixed coupling to mass ratios (denoted axion):
\begin{align}
&\,y_\mathrm{axion}\sim (x_\mathrm{axion}-4.66)-5
\nonumber \\
\to &\,
\frac{G^2_\mathrm{A(axion)}}{m^2_\mathrm{A(axion)}}=10^{-10}\frac{G^2_\mathrm{A(QED)}}{m^2_\mathrm{A(QED)}}
\;.
\end{align}
This is in agreement with the $\sim10^{-10}$ suppression of virtual pion vs QED effect in section~\ref{pionaxion}, recalling the similar (differences are model-dependent) products of decay constants and masses of axions and pions.

%%%%%%%%%%%%%%%%%%%%%%%%%%%%%%%%%%%%%%%%%%%%%
\subsection{PVLAS Result}

The summary of observational data presented in Figure 111.1 in~Ref.\,\cite{Tanabashi:2018oca} and copied into our Figure~\ref{fig2} includes an update from PVLAS, measuring vacuum birefringence and dichroism in external magnetic fields. While PVLAS 2006 results have been revised, with more recent results constraining $G_\mathrm{A}$ to a factor $\sim10$ smaller  compared to the 2006 result~\cite{Zavattini:2005tm,Zavattini:2007ee,DellaValle:2015xxa}, both results lie within the \lq virtual effect\rq\ region in Figure~\ref{fig2}, where an ALP virtual contribution surpasses the EHS effect. 

We find that the updated PVLAS range plotted, see \cite{DellaValle:2015xxa}, is included within the shaded region denoting prominent virtual effects:
\begin{align}
\frac{G^2_{A(\mathrm{PVLAS})}}{m^2_{A(\mathrm{PVLAS})}}
\sim&\, \Big(
\frac{10^{-7}(\mathrm{GeV}^{-1})}{10^{-3}(\mathrm{eV})}
\Big)^2
\nonumber \\
=&\,
10^{1.32}\frac{G^2_\mathrm{A(QED)}}{m^2_\mathrm{A(QED)}}
\;,
\end{align}
using Eq.\,(\ref{forplot4}) in the last line. Thus the PVLAS constraint  suggests a possible virtual pseudoscalar effect as much as  20 times the strength of the QED-EHS effect. 

%%%%%%%%%%%%%%%%%%%%%%%%%%%%%%%%%%%%
\section{Conclusion}

The ALP vacuum effect as shown in  Figure~\ref{fig2} covers a large domain provided by the astronomical observation constraints. These results rely on propagation of real ALPs, either over large distances from astronomical sources, or length scales probed by resonant cavities and light-through-wall experiments, see~\cite{Jaeckel:2010ni,Tanabashi:2018oca} and references therein. 

In the virtual ALP experiments EM fields must  be strong, preferably as near as possible to the  critical field  for the electron loop: $E_c\equiv m_e^2c^3/e\hbar=1.3\cdot 10^{18}\;\mathrm{V/m}$, in order to probe the effects inherent to EHS action and virtual ALPs. Considering the possibly small ALP mass range in  Figure~\ref{fig2}, we further need  EM fields quasi-constant over the range of the Compton wavelength of the ALP, which we benchmark for $m_\mathrm{A}=10^{-3}$\,eV at $\lambdabar_c=\hbar/m_\mathrm{A} c=2\times10^{-4}$m.

The virtual effect region of Figure~\ref{fig2} includes the domain of $G_\mathrm{A}$, $m_\mathrm{A}$ obtained by PVLAS. The PVLAS experiment probes virtual ALP and $e^+e^-$ effects via vacuum birefringence in an external magnetic field~\cite{DellaValle:2015xxa}. The external field driving the 4-photon interaction in Figure~\ref{fig1} consists of  a magnetic field constant over $\lambdabar_c$, and a $1064$nm laser~\cite{DellaValle:2015xxa}. 

This wavelength is smaller by a factor $\sim 10^{-2}$ than $\lambdabar_c$ for $m_\mathrm{A}\sim10^{-3}$\,eV, matching the Compton wavelength for larger mass $m_\mathrm{A}\sim10^{-1}$\,eV. Constraints on this larger mass via study of vacuum effects on external EM fields await resolution~\cite{Jaeckel:2010ni}. However, we note that virtual behavior of periodic fields may correspond to that of constant fields~\cite{Tomaras:2000ag,Avan:2002dn}, thus it is possible that the PVLAS experiment~\cite{Zavattini:2005tm,Zavattini:2007ee,DellaValle:2015xxa} using a standing laser wave and  an external magnetic field probes virtual effects on the required length scale. 

We note ongoing  effort to study the very strong field environments with high intensity lasers at ELI~\cite{Gies:2008wv,Dunne:2008kc}, and the long lasting study of supercritical fields in relativistic heavy-ion collisions~\cite{Ruffini:2009hg,Rafelski:2016ixr}. Both methods offer encouraging prospects as probing methods for QED + ALP action though the length and mass scale is today quite different, with EM fields  varying over shorter range than the usually applied $\lambdabar_c$ range. 

We conclude that our consideration of virtual ALP effects adds a new method for detection of ALPs that does not rely on propagation of real ALPs. Should ALPs exist only virtually in the vacuum (like quarks and gluons), they will never be discovered as free-streaming particles, but could via vacuum fluctuations we evaluated in this work. This observation adds to the findings we presented in  Figure~\ref{fig2} an additional motivation to  relevant experiments, such as PVLAS.

%%%%%%%%%%%%%%%%%%%%%%%%%%%%%%%%

\end{document}